\documentclass[aps,twocolumn,prl,superscriptaddress,showpacs]{revtex4}

\usepackage{graphicx}

\begin{document}

\title{Fluctuations and vortex pattern ordering \\
in fully frustrated $XY$ model with honeycomb lattice}

\author{S. E. Korshunov}
\affiliation{L. D. Landau Institute for Theoretical Physics,
Kosygina 2, Moscow 119334, Russia}
\affiliation
{Laboratoire de Physique Th\'{e}orique et Hautes
\'{E}nergies, CNRS UMR 7589,  \\
Universit\'{e} Paris VI and VII,
4 place Jussieu, 75252 Paris Cedex 05, France}

\author{B. Dou\c{c}ot}
\affiliation{Laboratoire de Physique Th\'{e}orique et Hautes
\'{E}nergies, CNRS UMR 7589,  \\
Universit\'{e} Paris VI and VII,
4 place Jussieu, 75252 Paris Cedex 05, France}

\date{\today}

\begin{abstract}

The accidental degeneracy of various ground states of a fully
frustrated $XY$ model with a honeycomb lattice is shown to survive
even when the free energy of the harmonic fluctuations is taken
into account. The reason for that consists in the existence of a
hidden gauge symmetry between the Hamiltonians describing the
harmonic fluctuations in all these ground states.
{A particular vortex pattern is selected only when anharmonic
fluctuations are taken into account. However, the observation of
vortex ordering requires relatively large system size $L\gg
L_c\gtrsim 10^5$}.

\end{abstract}


\maketitle

A fully frustrated $XY$ model can be defined by the Hamiltonian
\begin{equation}
H=-J\sum_{(ij)}\cos(\varphi_{j}-\varphi_{i}-A_{ij})\;,
\label{1}
\end{equation}
where $J>0$ is the coupling constant, the fluctuating variables
$\varphi_{i}$ are defined on the sites $i$ of some regular
two-dimensional lattice, and the summation is performed over the pairs
of nearest neighbors $(ij)$ on this lattice. The non-fluctuating
(quenched) variables $A_{ij}\equiv -A_{ji}$ defined on lattice bonds
have to satisfy the constraint $\sum A_{ij}=\pi\,(\mbox{mod}\,2\pi)$
(where the summation is performed over the perimeter of a plaquette)
on all plaquettes of the lattice.

For two decades such models (on various lattices) have been
extensively studied \cite{K02L} in relation with experiments on
Josephson junction arrays \cite{ML}, in which $\varphi_{i}$ can be
associated with the phase of the superconducting order parameter
on the $i$-th superconducting grain, and $A_{ij}$ is related to
the vector potential of a perpendicular magnetic field, whose
magnitude corresponds to a half-integer number of superconducting
flux quanta per lattice plaquette. Planar magnets with odd number
of antiferromagnetic bonds per plaquette \cite{Vil} are also
described by fully frustrated $XY$ models. Recently, the active
interest in fully frustrated Josephson arrays has been related to
their possible application for creation of topologically protected
quantum bits \cite{IF,DIF}.

The ground states of the fully frustrated $XY$ models are
characterized by the combination of the continuous \mbox{$U(1)$}
degeneracy (related with the possibility of the simultaneous
rotation of all phases) and discrete degeneracy related with the
distribution of positive and negative half-vortices between the
lattice plaquettes. Since vortices of the same sign repel each
other, the energy is minimized when the vorticities of the
neighboring plaquettes are of the opposite sign. In the case of a
square lattice this requirement is fulfilled for all pairs of
neigboring plaquettes when the vortices of different signs form a
regular checkerboard pattern \cite{Vil}. Analogous pattern, in
which the vorticities of the neighboring plaquettes are always of
the opposite sign, can be constructed in the case of a triangular
lattice \cite{LC,MS}.

In the case of a honeycomb lattice it is impossible to construct
a configuration in which the vorticities are of the opposite sign
for all pairs of neighboring plaquettes.
As a consequence, the discrete degeneracy of the ground state turns
out to be much more developed \cite{TJ,ShS}, and can be described in
terms of formation of zero-energy domain walls in parallel to each
other \cite{K}, the residual entropy of the system being proportional
to its linear size. Quite remarkably, the comparison of the free
energies of weak fluctuations in two different periodic ground
states shows that in harmonic approximation they are exactly equal to
each other \cite{K}, although the spectra of fluctuations
$\lambda({\bf k})$ in these states are essentially different:
for example, for small momenta ${\bf k}$ they are characterized by
different values of (anisotropic) helicity moduli.

In the present work we demonstrate that this absence of degeneracy
lifting is not a simple coincidence, but a consequence of a hidden
gauge symmetry between the Hamiltonians of harmonic fluctuations
in different ground states, and extends itself to all states
formed by some sequence of parallel zero-energy domain walls.
Therefore, the inclusion of harmonic fluctuations will not lead to
removal of the accidental degeneracy even if instead of
considering thermodynamic fluctuations one calculates the free
energy of quantum fluctuations at arbitrary (or zero) temperature.

This gauge symmetry is broken when anharmonicities are
taken into account. We have compared the leading anharmonic
contributions to the free energies corresponding to different ground
states and have found which of these states is selected at low
temperatures. However, the difference between their free energies
turned out to contain extremely small numerical coefficient.
As a consequence, the observation of vortex ordering 
is possible only in relatively large systems 
(whose size \makebox{$L\gg L_c\gtrsim 10^{5}$}).

We believe that the discovery of these new features
(some of which are completely unexpected) makes our results
of interest not only in relation with Josephson junction array
physics, but also in the more general context
of two-dimensional statistical mechanics.

$~$
\begin{figure}[b]
\includegraphics[width=2.0in]{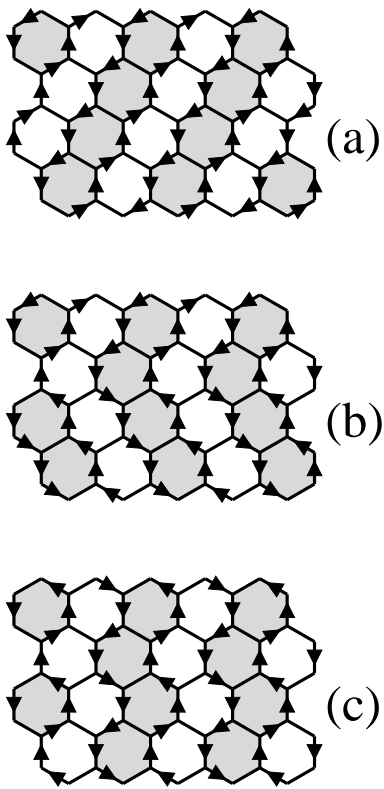}
\caption[Fig. 1] {Three different ground states of a fully
frustrated $XY$ model with a honeycomb lattice: (a) striped state,
(b) zero-energy domain wall and (c) zig-zag state. Directed arrows
(or their absence) correspond to $\theta_{ij}=\pi/4$
($\theta_{ij}=0$). Lattice plaquettes with positive vorticities
are shaded.} \label{fig1}
\end{figure}

Fig. 1a shows a structure of the simplest ground state of the fully
frustrated $XY$ model with a honeycomb lattice.
Each arrow corresponds to
$\theta_{ij}=\varphi_j-\varphi_i-A_{ij}=\pi/4$, whereas on the bonds
without arrows $\theta_{ij}=0$.
In this state the plaquettes with positive and negative vorticities
form straight stripes, so in the following we shall call it
a striped state.

A striped state allows for formation of domain walls
(separating two different realizations of such state)
which cost no energy \cite{K}.
An example of such zero-energy domain wall is shown in Fig. 1b.
An arbitrary number of zero-energy domain walls
separated by arbitrary distances can be introduced
into the system in parallel to each other \cite{K}.

If such domain walls are created at every possible position,
another periodic ground state is obtained, which is shown in Fig. 1c.
In accordance with the shape of the lines formed by the plaquettes
with positive and negative vorticities in this state
we shall call it a zig-zag state.
Alternatively, one can describe all other ground states
as obtained by the introduction of zero-energy domain walls
on the background of a zig-zag state.

If all sites of a honeycomb lattice are numbered by pairs of integers
$(n,m)$ as shown in Fig. 2, the Hamiltonian describing the harmonic
fluctuations in the striped state of Fig. 1a can be written as:
\begin{eqnarray}
H_a^{(2)} & = & \frac{1}{2}\sum_n\sum_{m=n(\mbox{\scriptsize
mod}\,2)}\left[ {J_1}(u_{n,m}-v_{n,m-1})^2+ \right. \label{2} \\
 & +& \left.{J_2}(u_{n,m}-v_{n+1,m})^2
+{J_3}(u_{n,m}-v_{n-1,m})^2 \right]        \nonumber
\end{eqnarray}
where $J_1=J_2=J\cos(\pi/4)$, $J_3=J$, whereas $u_{n,m}$ and $v_{n,m}$
are the deviations of the variables $\varphi$ from their equilibrium
values on two triangular sublattices forming a honeycomb lattice.

\begin{figure}[b]
\includegraphics[width=2.0in]{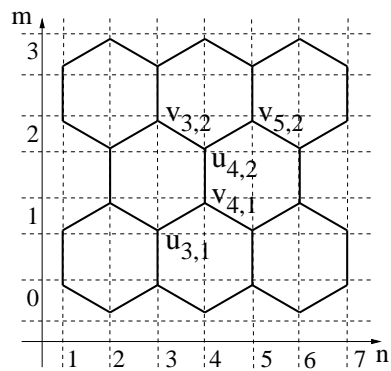}
\caption[Fig. 2] {The numbering of a honeycomb lattice sites by a
pair of integers $(n,m)$ used when writing Eq. (\ref{2}).}
\label{fxh-fig2}
\end{figure}

If one assumes the presence of periodic boundary conditions in the
horizontal direction and open boundary conditions in the perpendicular
(vertical) direction, the introduction of plane waves with respect
to the variable $n$ allows to rewrite Eq. (\ref{2}) as:
\begin{eqnarray}
H_a^{(2)} & = & \frac{1}{2}\int\frac{dq}{2\pi}\sum_{m}
\left\{J_S[|u_m(q)|^2+|v_m(q)|^2]-\right.  \label{3} \\
& - &\left.
J_1[u_{m+1}(q)v_{m}^*(q)+\mbox{c.c.}]-[K(q)u_m(q)v_m^*(q)+\mbox{c.c.}]
\right\}
\nonumber
\end{eqnarray}
where $J_S=J_1+J_2+J_3$ and
$K(q)=J_2\exp(-iq)+J_3\exp(iq)$.
A trivial gauge transformation:
\begin{equation}
\left(\begin{array}{l} u_{m+1}(q) \\
v_m(q) \end{array} \right) \Rightarrow  \exp[i\alpha(q)m]\times
\left(\begin{array}{l}\tilde{u}_{m+1}(q) \\
\tilde{v}_m(q)\end{array} \right),
\label{4}
\end{equation}
where $\alpha(q)=\arg[K(q)]$, allows then to replace Eq. (\ref{3}) by
the analogous expression with \makebox{${K}_0(q)\equiv|K(q)|$}
substituted for $K(q)$.

Note that in the chosen mixed representation the only modification of
the Hamiltonian which appears when an arbitrary sequence of
horizontal domain walls is introduced consists in replacement
of $K(q)$ by $K^*(q)$ for some values of $m$.
It is rather evident that the gauge transformation analogous
to Eq. (\ref{4}) (in which $\alpha(q)m$ should be replaced by
$\alpha(q)\sum_{m'<m}s_{m'}$, where variable $s_{m}=\pm 1$
describes the choice between the two options existing
for the continuation of a ground state at each $m$)
allows to transform any such Hamiltonian to the same form
(with $K(q)$ replaced everywhere by ${K}_0(q)$).

This means that for the boundary conditions described above,
the whole set of eigenvalues will be exactly the same
for all Hamiltonians obtained by the introduction of an arbitrary
sequence of horizontal domain walls (even for a finite sample).
Accordingly, the free energy of the harmonic fluctuations will be
exactly the same, and cannot be the source for the selection of
a particular ground state.

Clearly, the free energy of harmonic fluctuations also
remains degenerate when one considers a quantum generalization
of the same model with the diagonal mass term, which in terms of
a Josephson junction array corresponds to taking into account
the self-capacitance of each superconducting island \cite{FvdZ}.
In order to include into consideration the mutual capacitances of
neighboring islands (that is the capacitances of the junctions),
one has to apply the same approach (construction of
the gauge transformation which makes all the coefficients real)
not to the harmonic part of the Hamiltonian, but to the
frequency dependent Fourier transform of the harmonic contribution
to Euclidean Lagrangian, which also turns out to be possible.
That means that in the quantum version of the model the accidental
degeneracy survives (at the harmonic level) for arbitrary relation
between the self-capacitance of an island and the capacitance of
a junction.

If one assumes now periodic boundary conditions in the vertical
direction (instead of open boundaries), the degeneracy
of the free energy associated with harmonic fluctuations
(quantum or thermodynamic) will be manifested only
in the thermodynamic limit.

The accidental degeneracy of different ground states is removed
when anharmonicities are taken into account.
The leading contribution to the free energy induced by
anharmonic fluctuations, $F_{\rm anh}$,
is given by the sum of two terms,
which in the classical limit can be written as
\makebox{$F^{(3)}=-\langle[H^{(3)}]^2\rangle/2T$} and
$F^{(4)}=\langle H^{(4)}\rangle$, where $H^{(3)}$ and $H^{(4)}$
are, respectively, the third- and the fourth-order corrections
to the harmonic part of the Hamiltonian.

Since each term in $F^{(4)}$ depends only on local phase difference
on a particular bond, it can be proven with the help of
the hidden gauge symmetry discussed above that in the considered
system $F^{(4)}$ is the same for all ground states.
However, this property does not extend itself to $F^{(3)}$, which
depends also on more distant correlations.

Comparison of the expressions for $F^{(3)}$ in the two different
periodic ground states shows that the main contribution to
$\delta F_{\rm }=F^{(3)}_{\rm zig-zag}-F^{(3)}_{\rm str}$
(normalized per single hexagon) can be written as
\begin{equation}
\delta F_{\rm }=-6(G_s^3+G_z^3)J_*^2/T\;,           \label{dF}
\end{equation}
where $J_*=J\sin(\pi/4)/6$, whereas
\begin{equation}                                  \label{Gs}
G_s=\langle(u_{n,m}-v_{n+1,m})(u_{n+1,m-1}-v_{n+2,m-1})\rangle
\end{equation}
and
\begin{equation}                                  \label{Gz}
G_z=\langle(u_{n,m}-v_{n+1,m})(u_{n+1,m-1}-v_{n,m-1})\rangle
\end{equation}
are two correlation functions (for the bonds with
$|\theta_{ij}|=\pi/4$), calculated  with the help of the harmonic
Hamiltonian in striped and zig-zag states respectively.

Numerical calculation gives $G_s\approx 0.1559\,T/J$ and
$G_z\approx -0.1686\,T/J$, which after substitution in Eq. (\ref{dF})
leads to
$\delta F_{\rm }\approx \gamma T^2/J$, where
$\gamma\approx 0.8\cdot 10^{-4}$.
Addition to Eq. (\ref{dF}) of the terms associated with more distant
pairs of bonds leads only to a slight reduction of the
numerical coefficient to $\gamma\approx 0.7\cdot 10^{-4}$.

Thus we have demonstrated that $\delta F\propto T^2$, as in the case
of the antiferromagnetic $XY$-model with a {\em kagom\'{e}} lattice
\cite{HR,SKkagome}.
In situations, when so-called "order-from-disorder"
mechanism \cite{VBCC,Sh} works already at the harmonic level,
the free energy difference between the accidentally degenerate ground
states of frustrated $XY$ models is proportional to the first power of
$T$\cite{Kaw,K85,Henl,KVB,FCKF}.

The fluctuation induced free energy (per unit length) of a zero-energy
domain wall on the background of a striped state (see Fig. 1b)
is also given by $\delta F$.
It has been shown in Ref. \onlinecite{KVB} that in the frustrated
$XY$-models with accidental degeneracy which manifests itself
in the possibility of formation of zero-energy domain walls,
the temperature of the phase transition associated with
vortex pattern disordering (that is with the proliferation of such
walls) can be estimated as
\begin{equation}                                 \label{Tc}
T_c\approx E_K/\ln[T_c/\delta F(T_c)]\;,
\end{equation}
where $E_K\sim J$ is the energy of a kink on a domain wall and
$\delta F(T)$ is the fluctuation induced free energy of a domain wall
(per unit length).
For $\gamma\sim 10^{-4}$ the logarithmical factor in Eq.
(\ref{Tc}) is close to $12$, which means that the extreme smallness of
$\delta F(T)$ in the considered problem leads to the reduction
of $T_c$ (in comparison with $E_K$) by one order of magnitude.

However,
the extreme smallness of $\delta F(T)$ manifests itself much more
strongly in the relative prominence of the finite size effects.
For $\delta F(T)\approx \gamma T^2/J$ the (normalized) probability to
have a domain wall crossing a finite system of the width $L$ can be
estimated as $\exp[-(\gamma T/J)L]$ and is much smaller than one
only for $L\gg L_c=J/\gamma T$, which in our case gives
$L_c\gtrsim 10^5$.

This means that although anharmonic fluctuations in the fully
frustrated $XY$ model with a honeycomb lattice lead to the existence
of a thermodynamic phase transition (related with vortex pattern
ordering) at not too small temperature $T_c\sim 10^{-1}J$, the sizes
of Josephson junction arrays available experimentally, as well as the
sizes of the systems which can be simulated in numerical experiments
are currently not sufficient for observation of this ordering.

To conclude, in the present work we have demonstrated that the
fully-frustrated $XY$ model with a honeycomb lattice is the first
example of a statistical model in which the accidental degeneracy
of different ground states is not removed by the free energy of
harmonic fluctuations (neither in the classical, nor in the
quantum version of the model), although in different states these
fluctuations are described by different Hamiltonians. The
responsibility for that can be traced to a hidden gauge symmetry,
which manifests itself when these Hamiltonians are rewritten in
terms of the mixed representation.

We also have shown that accidental degeneracy is removed by the
free energy of anharmonic fluctuations, which leads to the selection
of a striped state of the type shown in Fig. 1a.
The state with analogous vortex configuration is selected as well in
a fully-frustrated superconducting wire network with honeycomb geometry
(in the vicinity of the phase transition),
but for entirely different reasons related with the possibility
of the order parameter modulation \cite{Xiao}.

However, the estimates based on numerical calculation of
the anharmonicity induced domain wall free energy
show that the system size ($L\gg 10^{5}$),
which is required for the observation of vortex ordering
in a fully frustrated Josephson junction array with a honeycomb
lattice is much larger than those which are typical for experiments
or numerical simulations, which makes the observation of such an
ordering rather problematic. It should be emphasized that although
the fully frustrated $XY$ model with a honeycomb lattice
has been the subject of Monte-Carlo simulations
of Shih and Stroud \cite{ShS,ShS2}, these authors have not analyzed
the structure of vortex pattern.

The ideas developed here may also be extended to other
geometries under current investigations. One of the most
intriguing systems in this respect is the fully frustrated
$XY$ model on a dice lattice, which exhibits a similar
degeneracy between its classical ground states \cite{SKdice},
and has been the subject of recent experiments \cite{Pannetier}
and numerical simulations \cite{CF}.
In particular, one of the main reasons for the absence of
vortex ordering in magnetic decoration experiments on
Josephson junction arrays \cite{Pannetier}, as well as in numerical
simulations of Ref. \onlinecite{CF} is very likely to be
a not sufficent system size.

The conclusion on relative prominence of finite size effects
(leading to the destructiion of long range order) in
situations when the stability of a vortex pattern is induced
only by anharmonic fluctuations is applicable to even
wider class of models. Their number includes
the antiferromagnetic $XY$ model with a {\em kagom\'{e}} lattice
\cite{HR,SKkagome,Rz}, in which, in the thermodynamic limit,
the anharmonic corrections to free energy lead
to the stabilization of so-called $\sqrt{3}\times\sqrt{3}$ state
at $T<T_c\sim 10^{-4} J$ \cite{SKkagome}. However, the estimates
analogous to those performed above allow one to conclude that
the observation of such an ordering is possible only when
$L\gg L_c\sim 10^7$, that is only in truely macroscopic system.
In numerical simulations of Ref. \onlinecite{Rz} this condition
definitely was not satisfied.

\vspace{1mm}

The work of S.E.K. has been supported in part by the Program
"Quantum Macrophysics" of the Russian Academy of Sciences and
by the Program "Scientific Schools of the Russian Federation"
(grant No. 00-15-96747).

\end{document}